\documentclass[aps,PRB,reprint,superscriptaddress,preprintnumbers]{revtex4-2}
\usepackage[T1]{fontenc}
\usepackage{times}
\usepackage{graphicx,color}
\usepackage{amsfonts,amsmath,amssymb,amsbsy}
\usepackage[colorlinks=true, linkcolor=blue, citecolor=blue, urlcolor=blue]{hyperref}
\usepackage{float}

\let\mathbf=\boldsymbol

\def\emph#1{\textcolor{blue}{#1}}
\begin{document}

\title{Mutual conversion between a magnetic N{\'e}el hopfion and a N{\'e}el toron}

\author{Shuang Li}
\affiliation{School of Science and Engineering, The Chinese University of Hong Kong, Shenzhen, Guangdong 518172, China}

\author{Jing Xia}
\affiliation{College of Physics and Electronic Engineering, Sichuan Normal University, Chengdu 610068, China}

\author{Laichuan Shen}
\affiliation{The Center for Advanced Quantum Studies and Department of Physics, Beijing Normal University, Beijing 100875, China}

\author{Xichao Zhang}
\affiliation{Department of Electrical and Computer Engineering, Shinshu University, Wakasato 4-17-1, Nagano 380-8553, Japan}

\author{Motohiko Ezawa}
\email[Email:~]{ezawa@ap.t.u-tokyo.ac.jp}
\affiliation{Department of Applied Physics, The University of Tokyo, 7-3-1 Hongo, Tokyo 113-8656, Japan}

\author{Yan Zhou}
\email[Email:~]{zhouyan@cuhk.edu.cn}
\affiliation{School of Science and Engineering, The Chinese University of Hong Kong, Shenzhen, Guangdong 518172, China}

\begin{abstract}
Three-dimensional (3D) magnetic textures attract great attention from researchers due to their fascinating structures and dynamic behaviors. Magnetic hopfion is a prominent example of 3D magnetic textures.
Here, we numerically study the mutual conversion between a N{\'e}el-type hopfion and a N{\'e}el-type toron under an external magnetic field. We also investigate the excitation modes of hopfions and torons in a film with strong perpendicular magnetic anisotropy.
It is found that the N{\'e}el-type hopfion could be a stable state in the absence of the external magnetic field, and its diameter varies with the out-of-plane magnetic field. The N{\'e}el-type hopfion may transform to a N{\'e}el-type toron at an out-of-plane magnetic field of about 20 mT, where the cross section structure is a N{\'e}el-type skyrmion. The hopfion and toron show different excitation modes in the presence of an in-plane microwave magnetic field.
Our results provide a method to realize the conversion between a N{\'e}el-type hopfion and a N{\'e}el-type toron, which also gives a way to distinguish Bloch-type and N{\'e}el-type hopfions.
\end{abstract}

\date{\today}
\keywords{Magnetic hopfion, toron, excitation mode, dynamics}
\pacs{75.10.Hk, 75.70.Kw, 75.78.-n, 12.39.Dc}

\maketitle

\section{Introduction}
\label{se:Introduction}

Topological solitons are stable particle-like configurations originally proposed in continuous-field theory~\cite{manton2004topological}.
The past few decades witnessed a continuous growth about topological spin textures in magnetic systems. 
A nanoscale example in one space dimension is the chiral domain wall~\cite{ref2,ref3,ref4}.
Magnetic skyrmions~\cite{skyrme1962unified,zhang2020skyrmion,sampaio2013nucleation,nagaosa2013topological,jiang2015blowing,yu2012skyrmion,nature442,properties6,anisotropy9} and magnetic vortices~\cite{van2006magnetic,doi:10.1063/1.4935276} are examples of two-dimensional (2D) topological solitons. 
Three-dimensional (3D) topological spin textures in magnetic materials have been observed recently including skyrmion strings~\cite{seki2021direct}, target skyrmions~\cite{zheng2017direct}, chiral bobbers~\cite{zheng2018experimental}, and hopfions~\cite{kent2021creation}.
Among these topological solitons, the 2D magnetic skyrmions in magnetic materials with Dzyaloshinskii-Moriya interactions (DMIs) have aroused much interest~\cite{1958A,1960Anisotropic} due to their prominent properties, such as their nanoscale size and low driving current density~\cite{2013Nucleation}. Besides, magnetic skyrmions can be used as information carrier in future spintronic applications~\cite{2013Skyrmions,0Nanoscale,2016Observation,2016Magnetic,2020A,2019Antiferromagnetic}. 

The magnetic skyrmion has a natural generalization to a topological soliton with a 3D topology, known as a magnetic hopfion. Similar with the clarifications of skyrmions, hopfions are identified with two types, that is, the Bloch-type and N{\'e}el-type hopfions. The cross section structure of a Bloch-type hopfions is a Bloch-type skyrmionium~cite{2015skyrmionic}, while the cross section structure of a N{\'e}el-type hopfion is a N{\'e}el-type skyrmionium~\cite{2019Current}.
The hopfion was originally proposed in the Skyrme-Faddeev model~\cite{1976Some}, where the topological number is not a generalized winding number but instead is a linking number of field lines, given by the integer-valued Hopf invariant. 
Several works have numerically predicted stable hopfions in the chiral magnetic system with the Dzyaloshinskii-Moriya interaction and interfacial perpendicular magnetic anisotropy (PMA)~\cite{2018Hopfions,2018Binding,2019Current} and reported that a Bloch hopfion can transform to a monopole-antimonopole pair (MAP)~\cite{2018Binding} or a toron~\cite{2021Field}.
Most recently, magnetic hopfions were experimentally created by adapting the PMA in Ir/Co/Pt multilayered systems~\cite{2020Creation}.

The basic element of hopfions and torons is a double twist torus~\cite{smalyukh2010three,chen2013generating}. Torons were initially observed in liquid crystals and also found in chiral magnets later~\cite{voinescu2020hopf,leonov2021surface,tai2020surface,rybakov2015new}. The transition between skyrmions and torons has been recently observed~\cite{tai2018topological,tai2020surface}. A magnetic toron is a spatially localized 3D spin texture composed of Bloch-type skyrmion layers with two Bloch points at its two ends~\cite{smalyukh2010three,leonov2018homogeneous}. Monopoles or Bloch points are topologically nontrivial point singularities, where magnetization suffers a discontinuity~\cite{malozemoff2016magnetic}.

The research of various dynamics of topological solitons is vital part in real applications and the studies of dynamic excitations could guide further design of microwave devices. 
The high-frequency gyrotropic modes of vortices have been studied during the past decades and the excitations of magnetic vortex can be applied in spin-torque oscillators~\cite{Youn2009Quantitative,2010Large,2011Dynamics} and spin-wave emitters~\cite{2016Magnetic}.
Spin excitation of skyrmions may offer more promising prospects toward the design of novel devices. 
Three characteristic resonant modes, which are breathing, clockwise (CW) and counterclockwise (CCW) modes, have been predicted theoretically in the year 2012~\cite{Mochizuki2011Spin}, and later on, these three modes were observed experimentally in the helimagnetic insulator using microwave absorption spectroscopy~\cite{2012Observation}.
Apart from the three aforementioned resonance modes, theoretical studies predict the existence of more collective excitations of skyrmions which have not been observed in experiments up until now~\cite{2014Internal,Sch2014Magnon,2017dynamics}.

Although numerous studies have been made on the dynamic excitation of low-dimensional topological solitons including vortices and magnetic skyrmions, 3D topological solitons have still not been well explored in this yield. The collective spin wave modes of the target skyrmion with the external field have been investigated~\cite{2019Collective,2017dynamics}. 
Recently, the resonant spin wave modes of the Bloch-type hopfion were studied by micromagnetic simulations~\cite{2021Field}, while the excitation modes of the N{\'e}el-type hopfion have not been reported.
%
\begin{figure}[t]
\centerline{\includegraphics[width=0.45\textwidth]{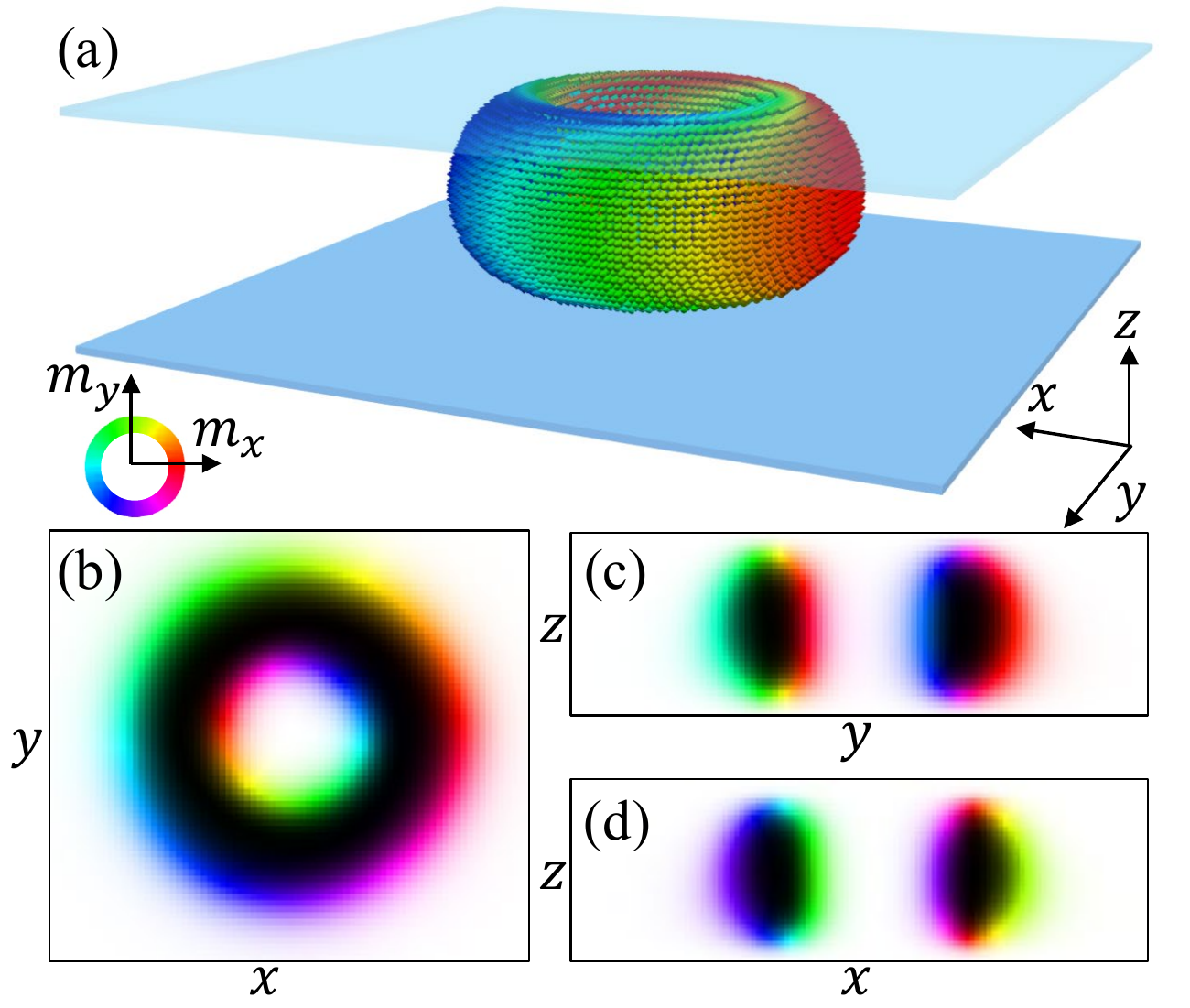}}
\caption{%
Schematic of the proposed structure.
(a) The thin cuboids at the top and bottom represent the magnetic films with strong PMA. The transparent region in the middle is the chiral magnet cuboid. The ring at the center represents the N{\'e}el-type hopfion. 
(b), (c), (d) Midplane cross-sectional spin configurations in the $xy$ plane $yz$ plane, and $xz$ plane. The color sphere and the coordinate system are shown in the insets.
}
\label{FIG1}
\end{figure}

In this work, we study the N{\'e}el-type hopfion and show that it is stable in a magnetic film in the absence of an external magnetic field. The film is sandwiched by two layers with high PMA. We also study another nontrivial state, the N{\'e}el-type toron consisting of N{\'e}el-type skyrmion layers with two point singularities at two ends, is also stabilized in this structure. We show that the N{\'e}el-type hopfion could transform to a N{\'e}el-type toron by an applied magnetic field.
Furthermore, we computationally study the resonant dynamics of the N{\'e}el-type hopfion and the N{\'e}el-type toron, which are excited by microwave magnetic field. The excitation modes are distinct between these topological spin textures. 
Especially, the resonant modes of N{\'e}el-type hopfions are different from that of Bloch-type hopfions~\cite{2021Field}.
We find a new type of 3D topological solitons called the N{\'e}el-type toron and our results offer a new observable way to clarify Bloch-type hopfions and N{\'e}el-type hopfions based on their excitation mode dynamics.
\section{Model and Methods}
\label{se:Model and Methods}

To stabilize a N{\'e}el-type hopfion in ferromagnetic material, we consider a magnetic film sandwiched by two PMA magnetic thin layers, as shown in Fig. 1(a). In the continuum approximation, the free energy is given by
\begin{equation}
    \begin{aligned}
    E=\int{[A_{\text{ex}}(\boldsymbol{\nabla{\rm{m}}})^2}+w_{\text{D}}+K_{\text{b}}(1-m_z^2)+\mu_0\\
{H_zM_s(1-m_z)]}{d}V+\int{K_{\text{s}}(1-m_z^2)}{d}S,
    \end{aligned}
\end{equation}
where $A_{\text{ex}}$ is the exchange constant. $w_\text{D}=D_i[m_{z}(\boldsymbol{\nabla}\cdot\boldsymbol{\rm{m}})-(\boldsymbol{\rm{m}}\cdot\boldsymbol{\nabla})m_{z}]$ is the interface-induced DMI energy density, which is favorable to hosting a stable N{\'e}el-type hopfion. $K_\text{b}$ and $K_\text{s}$ are the bulk and the interfacial PMA constant, respectively. $H_z$ is external magnetic field and $M_s$ is the saturation magnetization. The film has a length of $l=128$ nm and a height of $h=15$ nm, with each PMA layer on the top and bottom measuring $0.5$ nm. The meshed cell size is $0.5$ nm$\times$0.5 nm$\times$0.5 nm for total 2097152 cells per simulation. The non-local demagnetization energy is excluded for computational efficiency. 

We minimize the free energy (1) with an initial state computing from an ansatz~\cite{2019Current}. A $Q_\text{H}=1$ N{\'e}el-type hopfion is enabled in the structure in the absence of an external magnetic field and the midplane cross-sectional spin configurations are shown in Figs. 1(b)-(c) in three directions where the cross section structure in the $xy$ plane is a typical N{\'e}el-type skyrmionium and in the $yz$ or $xz$ plane is a skyrmion-antiskyrmion pair. The Hopf invariant is defined as
\begin{equation}
\label{eq:hopf} 
\ Q_\text{H}=-\int{\boldsymbol{\rm{B} \cdot \rm{A}}}{d^3}r,
\end{equation}
where $B_i=\frac{1}{8\pi}\epsilon_{ijk}\boldsymbol{\rm{m}}\cdot(\partial_j\boldsymbol{\rm{m}}\times\partial_k\boldsymbol{\rm{m}})$ is the emergent magnetic field calculating from spin texture, where $i,j,k=x,y,z$ and $\epsilon$ is the Levi-Civita tensor. And $\boldsymbol{\rm{A}}$ is a magnetic vector potential satisfying $\nabla \times \boldsymbol{\rm{A}}=\boldsymbol{\rm{B}}$~\cite{whitehead1947expression}.
Using this model, the relaxed states of the N{\'e}el-type hopfion under various external magnetic fields and the corresponding dynamic response to a microwave magnetic field are simulated by using the GPU accelerated micromagnetic simulation software package Mumax3~\cite{2014The}. Mumax3 performs a numerical time integration of the Landau-Lifshitz-Gilbert equation at zero temperature 
\begin{equation}
\label{eq:LLGn} 
\frac{d\boldsymbol{\rm{m}}}{dt}=-\gamma\boldsymbol{\rm{m}}\times \boldsymbol{\rm{H}}_{\rm{eff}}+\alpha\boldsymbol{\rm{m}}\times \frac{d\boldsymbol{\rm{m}}}{dt},
\end{equation}
where $\gamma$ is the gyromagnetic constant, $\alpha$ is the phenomenological Gilbert damping constant and $\boldsymbol{\rm{H}}_{\rm{eff}}$ is the effective field. 
Other parameters are based on previous numerical current-driven dynamics of hopfions in confined thin film~\cite{2019Current}. Explicitly, $A_{\text{ex}}=0.16$ pJm$^{-1}$, DMI strength of $D_i=0.115$ mJm$^{-2}$, $M_s=151$ kAm$^{-1}$, $K_\text{s}=1$ mJm$^{-2}$ and $K_\text{b}=20$ kJm$^{-3}$.

To study the dynamic response to magnetic field systematically, calculations in our simulations consist of two steps. Firstly, the relaxed states in the model are determined by static external magnetic fields along the $z$ axis with interval of 1 mT from $-20$ mT to 20 mT.
The energy is minimized to get an equilibrium state with a large damping constant of $\alpha=0.05$ for each value of the applied static field $H_z$.
Then the magnetization dynamics is computed about this equilibrium state by adding an additional \text{microwave} magnetic field in the plane. A sinc function, $h_x\frac{\sin(2\pi f_{\text{max}}t)}{2\pi f_{\text{max}}t}$, with a cutoff frequency of $f_{\text{max}}=15$ GHz and amplitude of $h_x=0.5$ mT, is used to excite the equilibrium states as its Fourier transform is a rectangular function. The simulation runs for $20$ ns with data sampled every $5$ ps, using a smaller value of damping constant, $\alpha=0.002$. The pulse is arbitrarily offset in time to peak at $1$ ps.
%

\section{Results and Discussions}
\label{se:Results and Discussions}
\begin{figure}[t]
\centerline{\includegraphics[width=0.48\textwidth]{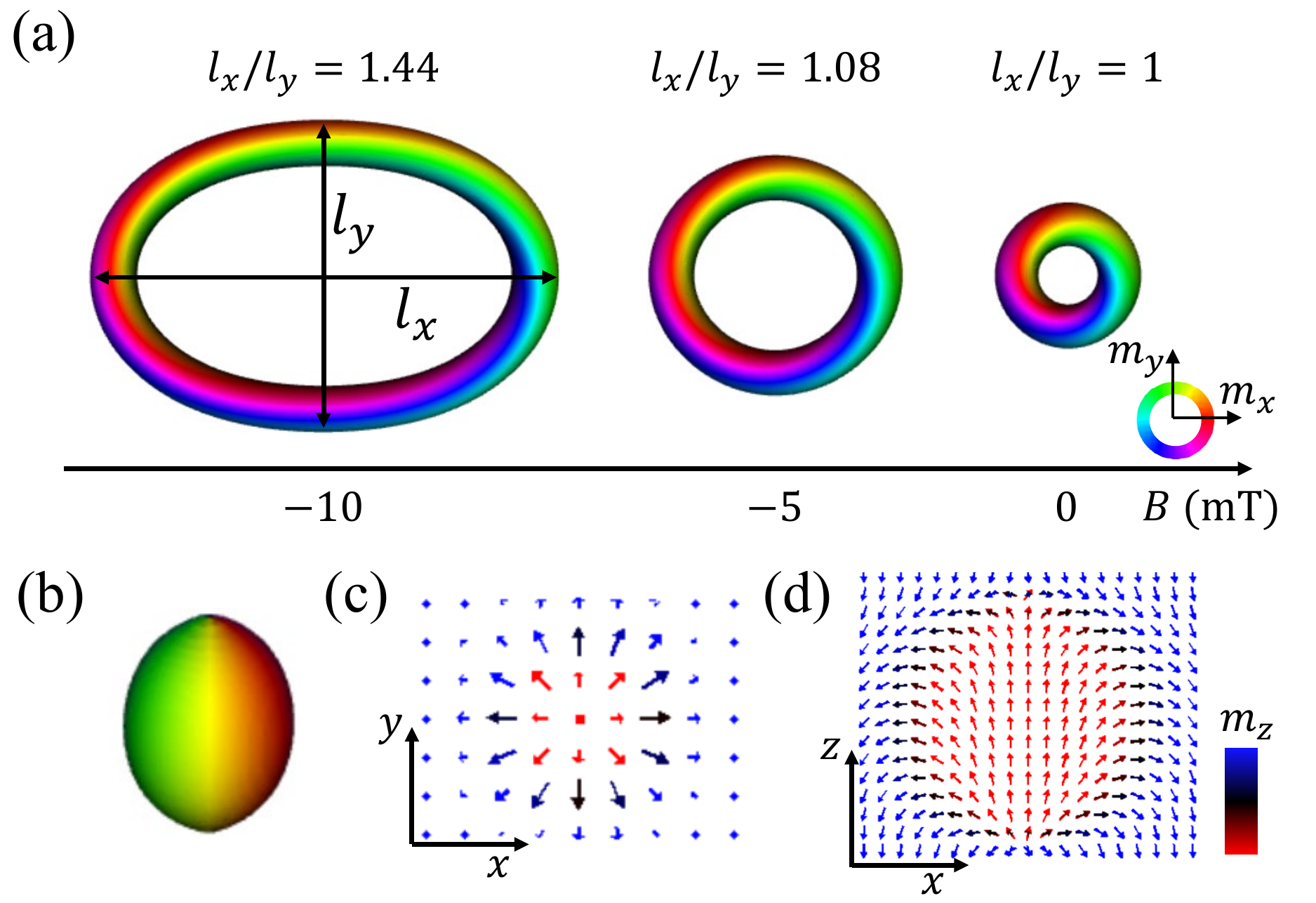}}
\caption{%
Spin configurations of enlarged N{\'e}el-type hopfions and a N{\'e}el-type toron.
(a) The set of preimages for the spin configurations of enlarged N{\'e}el-type hopfions with different aspect ratio with several external magnetic fields. $l_x$ and $l_y$ are the lengths of outer torus of N{\'e}el-type hopfion in the $x$ and $y$ directions. (b) The preimage of a N{\'e}el-type toron in the presence of out-of-plane magnetic field of 20 mT; (c) and (d) Midplane cross-sectional structures of a N{\'e}el-type toron in the $xy$ plane and $xz$ plane.
}
\label{FIG2}
\end{figure}

The spin configurations of the relaxed states are found to vary with the static external magnetic fields.
When a static magnetic field in the $-z$ direction is applied, the diameter of the N{\'e}el-type hopfion expands as the magnetic field increases. As a result, enlarged hopfions with different aspect ratios $l_x/l_y$ are obtained as shown in Fig. 2(a). $l_x$ and $l_y$ are the lengths of outer torus of the enlarged hopfion in the $x$ and $y$ directions, respectively. Obviously, the enlarged hopfion will deform when the magnetic field is large, resulting from the interaction with boundaries. The preimages of spin configurations are plotted using Spirit~\cite{muller2019spirit} for further visualization.
With applied field increasing in the $+z$ direction, the core spins of the N{\'e}el-type hopfion gradually shrink, and the cross section structure of the N{\'e}el-type hopfion transforms from the N{\'e}el-type skyrmionium~\cite{2017dynamics,2015skyrmionic} to the N{\'e}el-type skyrmion until 20 mT [Fig. 2(c)].
The preimage of a N{\'e}el-type toron is shown in Fig. 2(b). The N{\'e}el-type toron is similar with toron that previous works studied, while its cross section structure is not vortex-like but hedgehog-like. Thus, we name it N{\'e}el-type toron, and it is stable with wide range of external fields.
The cross section structure of a Bloch-type toron is a Bloch-type skyrmion, while the cross section structure of a N{\'e}el-type toron is a N{\'e}el-type skyrmion. They are both ends with two point singularities [Fig. 2(d)]. The topological charge for singularities is defined as~\cite{ostlund1981interactions}
\begin{equation}
\label{eq:q} 
q=\frac{1}{8\pi}\int{dS_i}\epsilon_{ijk}\boldsymbol{\rm{m}}\cdot \partial_j{\boldsymbol{\rm{m}}}\times \partial_k{\boldsymbol{\rm{m}}},
\end{equation}
Equation~(\ref{eq:q}) is also rewritten as $q=\frac{1}{4\pi}\int{\boldsymbol{\rm{G}}\cdot {d}{\boldsymbol{\rm{S}}}}$, where $\boldsymbol{\rm{G}}$ is gyrovector.
The topological charge density is defined according to the divergence theorem as
\begin{equation}
\label{eq:rho} 
\rho=\frac{1}{4\pi}\nabla \cdot \boldsymbol{\rm{G}}=\frac{1}{4\pi}{[\partial_xG_x+\partial_yG_y+\partial_zG_z]},
\end{equation}
where $G_i=\frac{1}{2}\epsilon_{ijk}{\boldsymbol{\rm{m}}}\cdot\partial_j\boldsymbol{\rm{m}}\times$$\partial_k{\boldsymbol{\rm{m}}}$.  For a toron, only the $\int_V{\partial_zG_z}dV$ is nonzero, and thus, the topological charge of a toron is $\pm1$.
%
\begin{figure}[t]
\centerline{\includegraphics[width=0.5\textwidth]{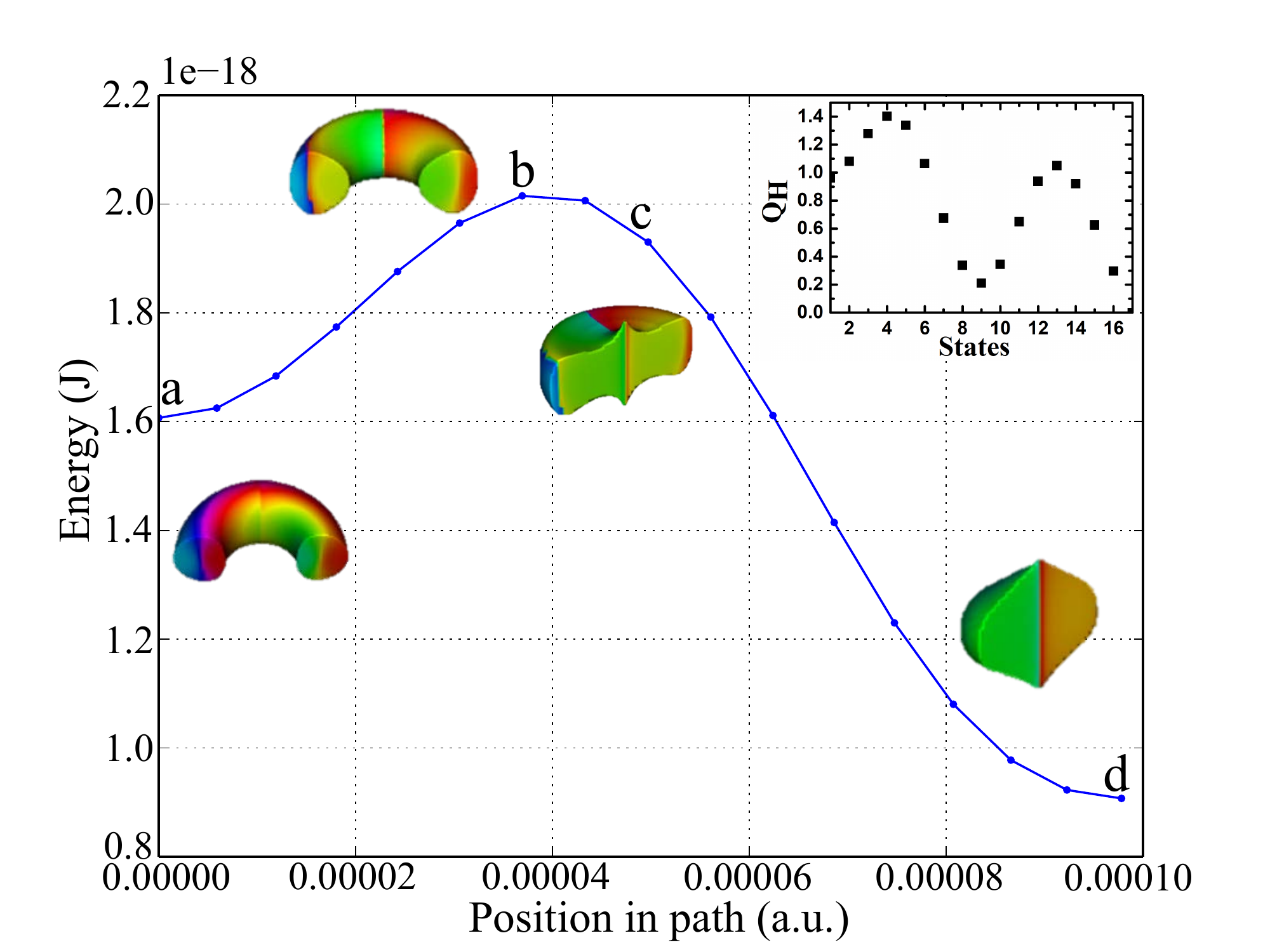}}
\caption{%
Minimal energy path between the N{\'e}el-type hopfion and the N{\'e}el-type toron.
The reaction coordinate is an order parameter that represents the relative distance between two neighboring states. Points a and d represent the N{\'e}el-type hopfion and the N{\'e}el-type toron, respectively. Point b is a saddle point, and the N{\'e}el-type toron is formed at point c. The inset is the Hopf invariant $Q_\text{H}$ from equation~(\ref{eq:hopf}) of each state along the minimal energy path.
}
\label{FIG3}
\end{figure}

\begin{figure*}[t]
\centerline{\includegraphics[width=1.0\textwidth]{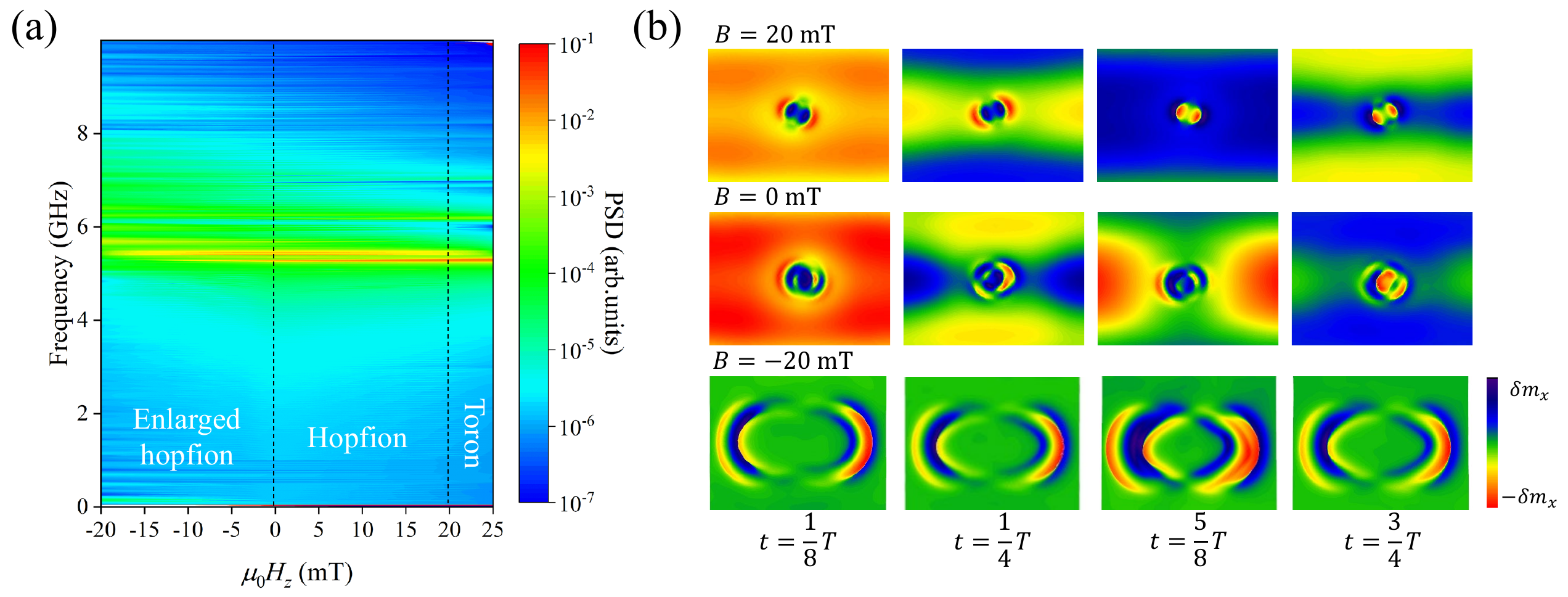}}
\caption{%
(a) Power spectral density (PSD) response of the spin texture as a function of the external static magnetic field. Two vertical dashed lines divide the spectrum area into three regions, each of which represents the enlarged hopfion, the N{\'e}el-type hopfion, and the N{\'e}el-type toron.
(b) Midplane snapshots of the real-time dynamics of $\delta{m_x}$ resolved at $-20$ mT (enlarged hopfion), 0 mT (N{\'e}el-type hopfion), and 20 mT (N{\'e}el-type toron). Each profile shows the results of averaging over the same instant of last ten periods.
}
\label{FIG4}
\end{figure*}

To further understand the topological conversion between the N{\'e}el-type hopfion and the N{\'e}el-type toron, a minimal energy path (MEP) calculation is performed between two states~\cite{bessarab2015method,bisotti2020fidimag}. The MEP calculation is carried out using the geodesic nudged elastic band (GNEB) method. Results from the MEP calculation are shown in Fig. 3. It exists an energy barrier between the N{\'e}el-type hopfion and the N{\'e}el-type toron, and an active energy is required to trigger the conversion between the N{\'e}el-type hopfion and the N{\'e}el-type toron. The initial state, the barrier peak, the intermediate state and the final state are plotted in Figs. 3(a)-3(d) for visualizing the conversion process.
Transformation from the N{\'e}el-type hopfion state (point a) to the intermediate state (point c) is caused by the slip of spins and reconnection of the preimages. And then, the intermediate state (point c) is relaxed to the N{\'e}el-type toron state (point d). Reverse process occurs with the magnetic field removed. In detail, the spins of central part rotate reversely, and also the point singularities move toward each other until they annihilate.
In the following, we discuss the difference of collective modes of the enlarged hopfion, the N{\'e}el-type hopfion and the N{\'e}el-type toron.
The dynamic resonances are calculated by exciting the equilibrium states with a time-varying field described by the sinc function in the presence of static perpendicular magnetic field. 
It is difficult to excite the N{\'e}el-type hopfion in this system by applying out-of-plane microwave magnetic field because of the strong boundary condition on the top and bottom layers. Furthermore, the finite thickness of this system matters, which is similar to the spin wave modes of 3D target skyrmions~\cite{2019Collective}. When the thickness is comparable with or even larger than the helical period, it is much easier to develop these collective modes along the disk normal direction due to the nature of spin waves and helimagnets. The thickness in this model is not comparable to the helical period of the material to get visible effect of the collective modes along the normal direction. Collective modes along the normal direction of the Bloch-type hopfion in a disk have been investigated~\cite{2021Field,2021spin}.
Therefore, only the resonances excited by in-plane microwave magnetic field are discussed here.
Simulations are performed in steps of 1 mT from $-20$ mT to 20 mT and the spatially average fluctuation in $x$ components of the magnetization are considered, $\delta{m_x(t)}=\langle m_{x,0} \rangle-\langle m_x(t) \rangle$,
where $m_{x,0}$ represents the $x$ components of the magnetization of the equilibrium state. The power spectrum density is computed from the Fourier transformation of this fluctuation.
In Fig. 4(a), we show calculated spectrum as a function of external perpendicular field $H_z$. Three spin textures are all resonant at around 5.4 GHz. For enlarged hopfion, a new peak occurs when the external magnetic field is larger than $-10$ mT, because the spin texture has slightly deformation. Over the critical transition field, that is 20 mT, the resonant frequency for toron increases slightly as $H_z$ increases. 

To further clarify the collective modes of each spin texture in details, the dynamics of the relaxed states is resolved with applying a stationary oscillating magnetic field, which is described by $h(t)=(h_x\sin(\omega_\text{R}t),0,0)$ with $h_x=0.5$ mT.
The resonance frequency $\omega_\text{R}$ is fixed at $5.4$ GHz for the common resonance frequency of three spin textures, and the period $T$ is $2\pi/\omega_\text{R}$.
Figure 4(b) shows the resulting profiles of fluctuations in the $x$ components of the magnetization. Each profile is obtained as follows.
Running with the uniform sinusoidal magnetic field applied along the $x$ axis over 20 periods $T$ of the excitation, the state is saved at time intervals of $T/8$ over the last ten periods of the simulations.
Thus eight snapshots of the state for each instant, $t_i=0$, 1/8$T$, 1/4$T$, $\cdots$, 7/8$T$, relative to the phase of the field excitation are obtained.
The final profile at each $t_i$ is then obtained by averaging over the ten snapshots, which allows artifacts to be averaged out over a single period of excitation~\cite{2014Breathing}. This method is called stroboscopic method. 

For enlarged hopfions as shown in Fig. 4(b), the relaxed state is obtained when a magnetic field of $-20$ mT is applied. The fluctuation of the $x$ components of magnetization shows expanding trend, which is similar with breathing mode of skyrmions. The more snapshots for detailed modes are shown in Ref.~\cite{SM}. The strong PMA layers confine the motion of the magnetization of the enlarged hopfion, and therefore there is no vertical mode. 
A N{\'e}el-type hopfion exists stably when magnetic field equals zero, and the resonant modes are collective modes. It shows the rotation of the outer torus and the expansion of the core of the N{\'e}el-type hopfion. At the same time, the spins of the torus of the hopfion flip reversely with the in-plane microwave magnetic field~\cite{SM}.
With 20 mT magnetic field applied, the situation is different for the N{\'e}el-type toron. It shows clearly the rotation mode with excitation, because the cross section structure is a typical N{\'e}el-type skyrmion~\cite{SM}. Relevant studies reported that the in-plane microwave magnetic field excites the rotation mode and the out-of-plane microwave magnetic field excites the breathing mode of skyrmions~\cite{2012Spin}. 
Unlike the resonant modes of the N{\'e}el-type hopfion, the Bloch-type hopfion show multiple resonant peaks~\cite{2021Field,2021spin}. And the modes of Bloch-type hopfion show hybridized modes with breathing and rotating characters.
\section{Conclusion}
\label{se:Conclusion}

In conclusion, we have studied numerically the N{\'e}el-type hopfion and the N{\'e}el-type toron in a film, and investigated their dynamic response to static magnetic field and microwave magnetic field.
This work shows the topological conversion between a N{\'e}el-type hopfion and a N{\'e}el-type toron with an active energy, which results from the reversal of spins and the reconnection of a torus.
Furthermore, dynamic simulations are employed to resolve individual resonant modes of the enlarged hopfion, the N{\'e}el-type hopfion and the N{\'e}el-type toron, and the power spectrum is calculated. The collective modes in 3D spin textures are hybridized modes due to 3D structures including breathing mode and rotation mode. 
Compared to the resonant modes in 2D skyrmions, breathing and rotation modes are not sufficient to characterize the modes in 3D topological solitons. Collective modes of enlarged hopfions, N{\'e}el-type hopfion and N{\'e}el-type toron show distinct behaviors.
We find a new type of 3D topological soliton called N{\'e}el-type toron. In addition, our results provide a method to realize the conversion between N{\'e}el-type hopfion and N{\'e}el-type toron, and it also gives evidence of clarifying the N{\'e}el-type hopfion and Bloch-type hopfion based on their different excitation modes.
%
\begin{acknowledgments}

This study is supported by Guangdong Special Support Project (Grant No. 2019BT02X030), Shenzhen Fundamental Research Fund (Grant No. JCYJ20210324120213037), Shenzhen Peacock Group Plan (Grant No. KQTD20180413181702403), Pearl River Recruitment Program of Talents (Grant No. 2017GC010293) and National Natural Science Foundation of China (Grant Nos. 11974298, 61961136006). 
J.X. acknowledges the support by the National Natural Science Foundation of China (Grant No. 12104327).
X.Z. was an International Research Fellow of the Japan Society for the Promotion of Science (JSPS). X.Z. was supported by JSPS KAKENHI (Grant No. JP20F20363).
M.E. acknowledges the support by the Grants-in-Aid for Scientific Research from JSPS KAKENHI (Grant Nos. JP17K05490 and JP18H03676) and the support by CREST, JST (Grant Nos. JPMJCR16F1 and JPMJCR20T2).
\end{acknowledgments}



\end{document}